\documentclass[twocolumn,showpacs,preprintnumbers,amsmath,amssymb,prl]{revtex4}

\usepackage{graphicx}
\usepackage{dcolumn}
\usepackage{bm}

\begin{document}

\preprint{APS/123-QED}

\title{Non-magnetic left-handed material}%

\author{Viktor A. Podolskiy}
\email{vpodolsk@princeton.edu}
\homepage{http://corall.ee.princeton.edu/vpodolsk}
\author{Evgenii E. Narimanov}
\affiliation{Electrical Engineering Department, Princeton University, Princeton, NJ 08544}

\date{\today}

\begin{abstract}
We develop a new approach to build a material with negative refraction index. In contrast to conventional designs which make use of a resonant behavior to achieve a non-zero magnetic response, our material is intrinsically non-magnetic and relies on an anisotropic dielectric constant to provide a left-handed response in waveguide geometry. We demonstrate that the proposed material can support surface (polariton) waves, and show the connection between polaritons and the enhancement of evanescent fields, also referred to as super-lensing. 
\end{abstract}

\maketitle
A large number of potential applications in optics, materials science, biology, and biophysics has instigated an extensive research in the area of materials with negative phase velocity, also known as left-handed media (LHM). Effects predicted in LHMs include reversed Snell's law, Doppler Effect, Cherenkov radiation~\cite{veselago}, super-lensing~\cite{pendry,pendryCyl}, multi-focal imaging~\cite{efros}, unnatural nonlinearities~\cite{kivshar,agranovich}, and soliton propagation~\cite{kivshar}. The LHMs, associated with simultaneously negative dielectric permittivity and magnetic permeability have been successfully demonstrated in {\it microwaves} in (i)~a composite of antennas (which provide dielectric response) and split-rings resonators (responsible for magnetic resonant properties)~\cite{smith,Parazzoli}, and (ii)~photonic crystals~\cite{sridhar}. However, most of practical applications of these unique materials are in the ``faster'' ({\it optical and infrared}) region of the spectrum. 

The direct scale-down of the experimentally verified split-ring designs is currently possible only to THz-frequencies~\cite{smithTHz}, and can hardly be extended farther. Other proposed designs of infrared and optical LHMs~\cite{JNOPM2002,podolskiyOptExp,shvetsPRB} are (i)~generally extremely sensitive to fabrication defects and (ii)~often rely on a resonance to provide negative magnetic response. The resonant losses, which accompany the hard-to-achieve negative magnetic permeability at high frequencies~\cite{landau} make the present LHM designs almost impractical for real-life applications. The only exception from this ``resonant'' approach so far is Ref.~\cite{shvetsPRB}, which again represents a great fabrication challenge. 

In this letter we propose to use a planar waveguide with anisotropic dielectric core as left-handed media. Unlike most of the present LHM composites, our system does not have any magnetic response. Moreover, in contrast to current {\it composite, resonance-based} LHM designs, the proposed material may be homogeneous, and does not require a resonance to achieve a negative phase velocity, the fundamental property of LHM. The proposed material may support surface (polariton) waves at the interface between LHM and right-handed media. We describe the electromagnetic properties of our system, and derive the conditions for its right-, and left-handed response, and for excitation of surface waves (polaritons). We show the connection between the existence of polaritons and enhancement of exponentially decaying (evanescent) fields. Finally, we consider the fabrication perspectives of the materials described in this paper and propose several feasible designs for {\it optical and infrared} LHMs. 

\begin{figure}
\centerline{\includegraphics[width=8cm]{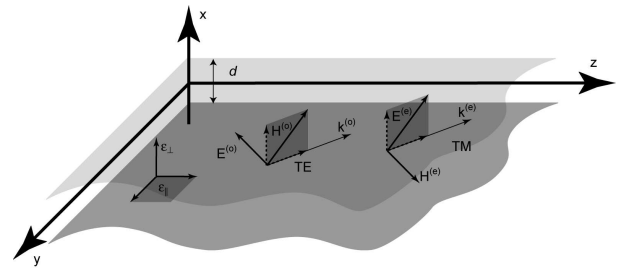}}
\caption{
\label{figConfig} The extraordinary (TM) and ordinary (TE) waves propagating in a planar waveguide with anisotropic core.}
\end{figure}

We consider a planar waveguide, parallel to the $(y,z)$ plane of coordinate system, with the boundaries at $x=\pm d/2$. We assume that the material inside the waveguide is non-magnetic ($\mu=1$), and has anisotropic {\it uniaxial} dielectric constant $\epsilon$, with $\epsilon_x=\epsilon_{\perp}$ and $\epsilon_y=\epsilon_z=\epsilon_{||}$ (see Fig.~\ref{figConfig}). 

Similarly to uniaxial crystals, our system may support two different kinds of electromagnetic waves \cite{landau}. The waves of the first kind have their {\it electric field vector in $(x,y)$ plane}. The propagation of such waves depends only on $\epsilon_{||}$, and is not affected by anisotropy. These waves are also known as {\it ordinary waves}. In contrast to this behavior, the waves of the second kind (known as {\it extraordinary waves}) have their {\it magnetic field in $(x,y)$ plane}. Correspondingly their electromagnetic properties are affected by both $\epsilon_{||}$ and $\epsilon_{\perp}$. 

As we show below, the ordinary and extraordinary waves are fundamentally distinct as they have different dispersion relations and refraction properties. 

A wave propagating in the proposed system can be represented as a series of the waves with their electric (magnetic) field perpendicular to the direction of propagation, known as TE (TM) waves correspondingly (see, e.g.~\cite{landau}). In our case of the planar waveguide with anisotropic core, extraordinary wave has TM polarization, while ordinary wave has the TE form (see Fig.~\ref{figConfig}). As it can be explicitly verified, the $\{x,y,z\}$ components of ordinary $(E^{(o)},H^{(o)})$, and extraordinary $(E^{(e)},H^{(e)})$ waves propagating in $(y,z)$ direction can be represented by the following expressions:
\begin{eqnarray}
\nonumber
E^{(e)}&=&\left\{i\frac{k^{(e)^2}_z+k^{(e)^2}_y}{k^{(e)}_z \varkappa^{(e)^2}}
\frac{\epsilon_{||}}{\epsilon_{\perp}} {E^{(e)}_0}' ;\;
\frac{k^{(e)}_y}{k^{(e)}_z}E^{(e)}_0;\;
E^{(e)}_0\right\}\\
\label{eqFields}
H^{(e)}&=&\left\{0; \;
i\frac{{ k} \epsilon_{||}}{\varkappa^{(e)^2}} {E^{(e)}_0}'; \;
-i\frac{{ k} k^{(e)}_y \epsilon_{||}}{k^{(e)}_z\varkappa^{(e)^2}} {E^{(e)}_0}'
\right\}
\\
\nonumber
E^{(o)}&=&\left\{0; \;
E^{(o)}_0;\;
-\frac{k^{(o)}_y}{k^{(o)}_z}E^{(o)}_0
\right\}
\\
\nonumber
H^{(o)}&=&\left\{
-\frac{k^{(o)^2}_z+k^{(o)^2}_y}{{k} k^{(o)}_z}E^{(o)}_0; \;
-\frac{i k^{(o)}_y}{{ k}k^{(o)}_z}{E^{(o)}_0}'; \;
-\frac{i}{ k}{E^{(o)}_0}'
\right\},
\end{eqnarray}
where ${ k}=\omega/c$, and prime ($'$) denotes the differentiation with respect to $x$. Similarly to Ref.~\cite{landau} the field $E^{(e|o)}_0(x,y,z;t)=E^{(e|o)}_0(x){ e}^{-i\omega t+i k^{(e|o)}_y y+i k^{(e|o)}_z z}$ is defined from the equation 
\begin{equation}
\label{eqE0}
{E^{(e|o)}_0}''+\varkappa^{(e|o)^2} E^{(e|o)}_0=0,
\end{equation}
with the boundary conditions satisfying the conditions for tangential ($y,z$) components of the electric field at the waveguide walls. For simplicity, here we consider the case of perfectly conducting waveguide boundaries; the straightforward extension of the presented theory to the case of dielectric walls, where similar effects are anticipated, will be presented elsewhere. 

The Eq.~(\ref{eqE0}) yields a series of solutions (modes) $E^{(e|o)}_0(x)=A^{(e|o)}_m \cos(\varkappa x)$ with $\varkappa=(2m+1)\pi/d$, and $E^{(e|o)}_0(x)=A^{(e|o)}_m \sin(\varkappa x)$, $\varkappa=2m\pi/d$ (where $n$ is an integer number). Each waveguide mode has its own dispersion relation: 
\begin{eqnarray}
\label{eqDispEq}
k^{(e|o)^2}_z+ k^{(e|o)^2}_y=\epsilon^{(e|o)}\nu^{(e|o)}{ k}^2,
\end{eqnarray}
where 
\begin{eqnarray}
\epsilon^{(e)}=\epsilon_{\perp};\; \epsilon^{(o)}=\epsilon_{||}; \;
\nu^{(e|o)}=\left(1-\frac{\varkappa^{(e|o)^2}}{\epsilon_{||}{ k}^2}\right)
\end{eqnarray}

Note that due to different geometry the TM and TE modes defined here are somewhat different from the conventional waveguide solutions presented in common textbooks \cite{landau}. Here we focus on the planar waveguide {\it unbounded in $(y,z)$ plane} with anisotropic core in contrast to {\it bounded in $(x,y)$ directions} ``tubular'' ($1D$) structure with isotropic filling, where waves can propagate in $z$ direction alone. Taking into account the identical dispersion relations for ordinary and extraordinary waves in the case $\epsilon_{\perp}=\epsilon_{||}$, it is straightforward to obtain the well-known TE ($E_z=0$) and TM ($H_z=0$) ``tubular'' solutions as the linear combination of the waves from Eqs.~(\ref{eqFields}). Also, as an alternative to the formalism presented in this letter, our system may be described in terms of introduced in \cite{agranovich} generalized dielectric tensor with spatial dispersion. 

An arbitrary wave inside a planar waveguide can be represented as a linear combination of waveguide modes (corresponding to different values of $\varkappa$). For simplicity for the rest of the letter we limit ourselves to the case when only a single mode is excited. This assumption does not restrict the generality of our approach since it does not limit $(y,z)$ structure of the solutions or their polarization. The generalization of expressions presented here, to a multiple-mode case is straightforward. 

\begin{figure}
\centerline{\includegraphics[width=8cm]{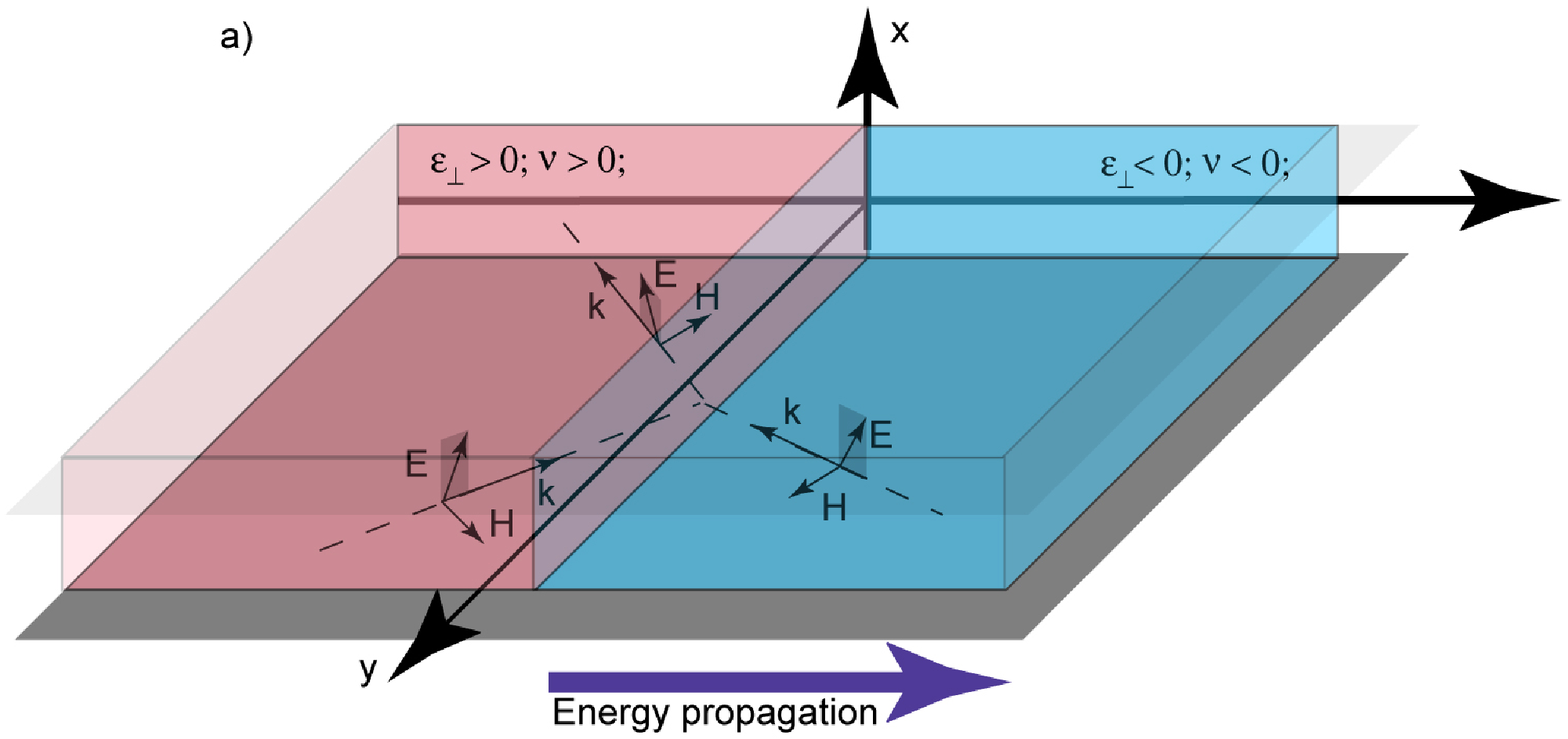}}
\centerline{\includegraphics[width=6cm]{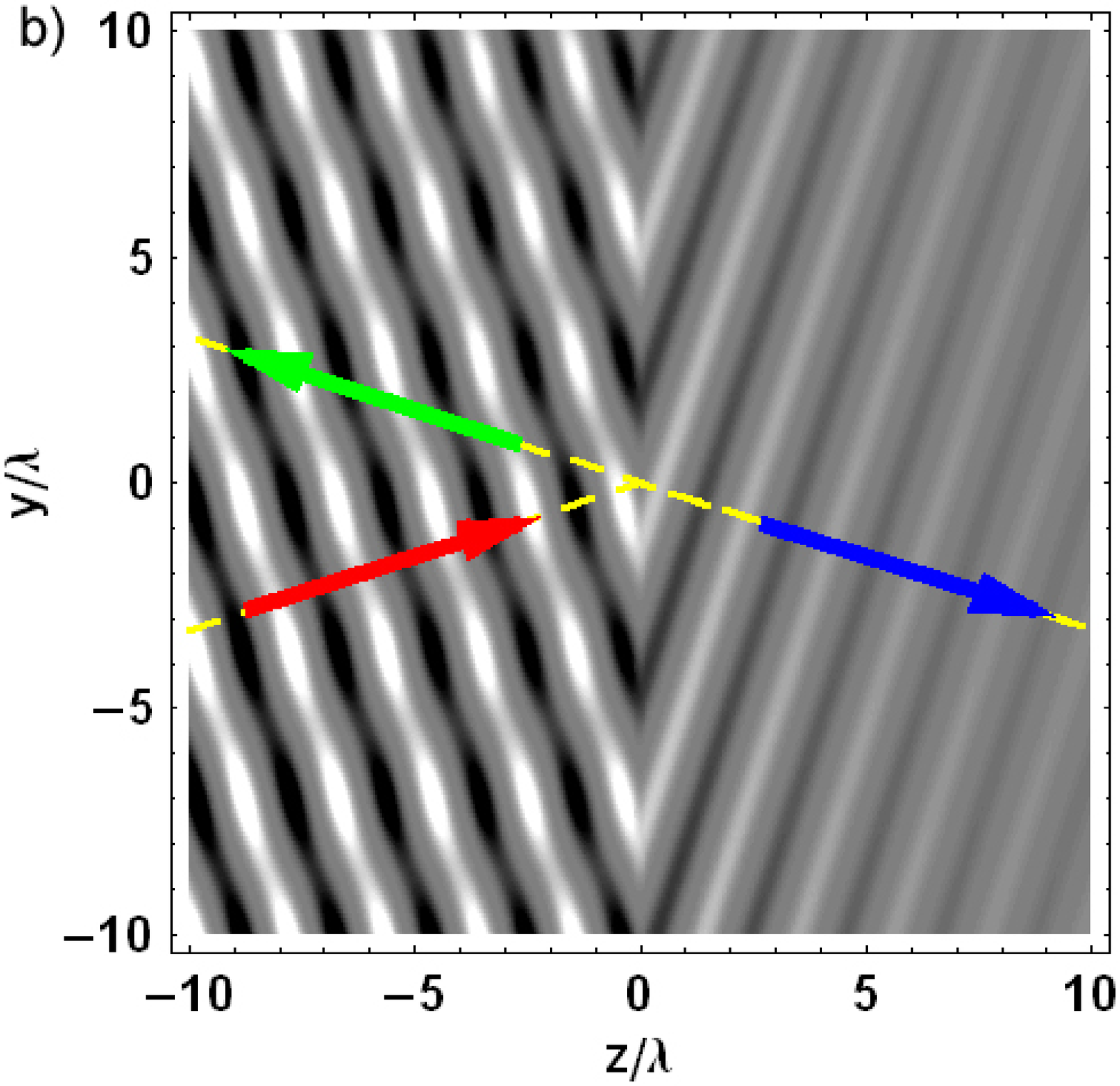}}
\caption{
\label{figRefract} Reflection and refraction on the boundary between isotropic (right-) and left-handed media. (a)~Schematic illustration of refraction of a TM wave at right- and left- handed media interface (ordinary wave not shown). (b)~Results of exact numerical calculations of refraction of the mode $\varkappa={k}/2$ in a planar waveguide. The dielectric parameters in right-handed media ($z<0$): $\epsilon=\nu=1/2$; in the left-handed media ($z>0$): $\epsilon=\nu=-1/2$, angle of incidence is $\pi/10$. Red, green, and blue, arrows show the direction of incident, reflected and refracted waves correspondingly}
\end{figure}

It is clearly seen from Eq.~(\ref{eqDispEq}) that a propagating solution (described by real $k_z$ and $k_y$) is only possible in the case when the corresponding parameters $\epsilon$ and $\nu$ are of the same sign. The case $\epsilon>0; \;\nu>0$ is usually realized for an isotropic material inside the planar (transmitting) waveguide~\cite{landau}; the case $\epsilon>0; \nu<0$ corresponds to the so-called subcritical waveguide which does not support propagating modes and reflects all ``incident'' radiation. The third case which can be realized in a waveguide with isotropic core, $\epsilon<0; \nu>0$, describes a perfectly conducting interior, which again does not support propagating waves. 

Finally the case $\epsilon<0; \nu<0$, which is a primary focus of this letter, can only be realized only for {\it the extraordinary wave in the anisotropic material}. The corresponding structure is transparent for TM wave; the TE solution exponentially decays into such a waveguide. 

While Eq.~(\ref{eqDispEq}) defines the magnitude of the phase velocity of the mode in this case, the {\it sign} of the phase velocity cannot be determined by Eq.~(\ref{eqDispEq}) alone. To define the sign of the phase velocity, and consequently the ``handedness'' of a media, we consider the refraction of a wave at the interface between the transparent isotropic (right-handed) media and a media with $\epsilon<0; \nu<0$ inside the same waveguide. We assume that the interface coincides with the coordinate plane $z=0$ (see Fig.~\ref{figRefract}). 

We first consider the special case of the normal propagation of a TM-polarized wave. Since in such a wave $H_z=H_x=0$, neither refracted nor reflected ordinary waves are excited. Since for $k_y=0$ the components $H_y$ and $E_x$ are related to each other: $H_y=\frac{{ k}\epsilon_{\perp}}{k_z}E_x$ [see Eqs.~(\ref{eqFields})], the requirement for continuity of tangential fields across the boundary $z=0$ immediately shows that the sign of $k_z$ should coincide with the one of $\epsilon_{\perp}$. This is a clear indication that the media with $\epsilon<0,\;\nu<0$ is left-handed. 

The analysis of a general case of obliquely incident wave (shown in Fig.~\ref{figRefract}) is more complicated, as in general ordinary reflected wave is also excited, and the direction of the refracted (extraordinary) wave should be determined by the causality principle \cite{landau}. We perform such an analysis via exact numerical calculations where we assert that the propagating in the real (absorbing) media wave decays in the direction of its propagation. The results of these calculations are shown in Fig.~\ref{figRefract}. It is clearly seen that Snell's law is reversed, meaning that phase velocity in the medium with $\epsilon<0; \nu<0$ is negative and the resulting wave is left-handed for a general case of oblique incidence. As it is shown in~\cite{veselago}, all optical effects directly related to a phase velocity (Snell's law, Doppler Effect, Cherenkov radiation etc.) are reversed in such a media. 

Another class of phenomena commonly associated with LHMs (e.g. enhancement of the evanescent fields~\cite{pendry}, nonlinear surface waves~\cite{kivshar}), however requires the propagation of surface waves, also known as polaritons, at the left- and right-handed media interface. As it is shown in \cite{landau}, only TM waves could construct a surface mode on a non-magnetic interface. In the following calculations we represent the fields and electromagnetic constant of right-handed media (which fills the region $z<0$) with superscript $(-)$ and the ones in LHM region $z>0$ with $(+)$. We search for a polariton solution $(E,H)^{(-)}\propto \exp[i k_y y+\xi^{(-)}z]; \;(E,H)^{(+)}\propto \exp[i k_y y-\xi^{(+)}z]$, with {\it real} $k_y$, and positive $\xi^{(-|+)}$ (the exponentially-growing away from the interface ``anti-polariton'' solution corresponding to negative $\xi^{(-|+)}$ can exist only in finite space region). 

While the LHM region is bound to have $\epsilon_{\perp}<0,\;\epsilon_{||}>0$, the ``right-handed'' medium can be constructed by either $\epsilon_{\perp}>0, \; \epsilon_{||}>0$ or by $\epsilon_{\perp}>0,\; \epsilon_{||}<0$. These two combinations of the dielectric constants lead to different conditions for polariton propagation. 

Specifically, for the case $\epsilon^{(-)}_{||}>0, \epsilon^{(-)}_{\perp}>0$, usually realizable in an isotropic right-handed medium, the polaritons are only possible for $k_y=0$ and have the dispersion relation (see Eqs.~(\ref{eqFields})):
\begin{eqnarray}
\label{eqPolarI}
\frac{\nu^{(-)}}{\epsilon^{(-)}_{\perp}}=
\frac{\nu^{(+)}}{\epsilon^{(+)}_{\perp}} 
\end{eqnarray}
Such waves however assume propagation along $x$ direction. The existence of these waves in the waveguide geometry considered here is limited to a number of ``modes'', each forming a standing wave between the waveguide plates and fulfilling the corresponding boundary conditions (see also Eq.(\ref{eqE0}) and the discussion afterwards). 

\begin{figure}
\centerline{\includegraphics[width=8cm]{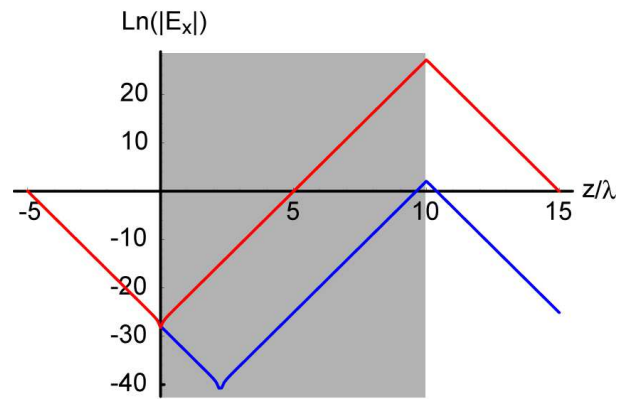}}
\caption{
\label{figPolariton}Amplification of evanescent field by a parallel slab of planar LHM ($\varkappa=k/2$). The blue line corresponds to non-plasmonic case. Right-hand media (RHM) parameters: $\epsilon^{(RHM)}=\nu^{(RHM)}=1/2$, LHM parameters: $\epsilon^{(LHM)}=\nu^{(LHM)}=-1/2$; $k_y=2 k$. Red line shows the case of resonant excitation of polariton waves $\epsilon^{(RHM)}=3/2, \nu^{(RHM)}=4; \epsilon^{(LHM)}=-6/5, \nu^{(LHM)}=-5; k_y=\sqrt{9/8}k$; the LHM is positioned between $z=0$ and $z=10\lambda$. The resonant enhancement of evanescent components with surface waves (often attributed to superlens, originally proposed in \cite{pendry}) is clearly seen.}
\end{figure}

However if the right-hand medium has $\epsilon_{||}<0$, and $\epsilon_{\perp}>0$, the propagation of polaritons with non-zero $k_y$ is also possible when 
\begin{eqnarray}
\label{eqPolarII}
\epsilon^{(-)}_{||} \nu^{(-)}=\epsilon^{(+)}_{||} \nu^{(+)}
\end{eqnarray}
This equation again relates $\varkappa$ to $k$. When Eq.~(\ref{eqPolarII}) is satisfied, the surface wave exists for any given $|k_y|^2>\epsilon\nu k^2$, and the relation between $k_y$ and $\xi$ is given by Eq.~(\ref{eqDispEq}), where we substitute $k_z^2=-\xi^2$. Note that a similar situation takes place in $3D$ geometry on the boundary between the right-handed medium ($\epsilon^{(-)}>0, \mu^{(-)}>0$) and ``conventional'' LHM ($\epsilon{(+)}<0,\mu{(+)}<0$), where for the same frequency the polaritons exist for any wavevector provided that $\epsilon^{(-)}=-\epsilon{(+)}, \mu^{(-)}=-\mu{(+)}$. 

We stress that it is the existence of surface waves for a wide range of wavevectors which makes the proposed in \cite{pendry} phenomenon of super-lensing possible. The evanescent components, which carry the information about the subwavelength features of the source, exponentially decay away from the object plane. Their resonant enhancement by a slab of either planar (described here) or $3D$ (described in Refs.~\onlinecite{pendry,veselago}) LHM can be represented as a resonant coupling of the original evanescent wave to the surface modes on {\it both interfaces of the LHM lens}. In such a process, the original evanescent wave excites anti-polariton (surface mode growing away from the interface) on the front interface (see Fig.~\ref{figPolariton}), which in turn excites the true-polartion mode on the back interface of the slab. The exponentially decaying away from the lens part of this surface mode represents the LHM-enhanced evanescent wave. This concept is illustrated in Fig.~\ref{figPolariton} where we calculate the transmission of an evanescent component through the slab of planar LHM proposed here. It is clearly seen that decaying through the right-handed media evanescent wave, is resonantly enhanced inside LHM slab only in the presence of polaritons~\cite{losses}. 

Finally, we consider the fabrication perspectives of the proposed LHM materials. In optical and near-infrared frequencies negative dielectric constant is easily achieved due to plasmon resonance of a free electron gas inside the metal~--~(Ag, Au, etc.) or doped semiconductor~--~(Si) (plasmonic) structures. Electron concentration and their effective mass plays a role of tuning parameter, which defines the exact position of plasmon resonance, and consequently the {\it highest $\epsilon<0$ frequency}. In mid-infrared spectrum range negative dielectric constant naturally occurs in polar crystals (e.g. SiC)~\cite{spitzer}. Both plasmonic and polar materials generally have small absorption, which in conjunction with negative $\epsilon$ makes them excellent candidates for LHM preparation. The anisotropic dielectric response described here, may be achieved in the following composites: 

(i)~Composite of subwavelength (nanostructured) inclusions with anisotropic (e.g. spheroidal) shape in isotropic dielectric host. In this approach all the inclusions have to be aligned, and homogeneously distributed in the dielectric host. The shape of the inclusion defines the frequency range of LHM response. We stress that no special arrangement of the inclusions (except for their aligning) is necessary to achieve a desired dielectric properties. 

(ii)~Composite based on isotropic (spherical) inclusions in a dielectric host. The anisotropy may be achieved by anisotropic concentration of inclusions. For example, one may deposit a dielectric spacer followed by (random) deposition of inclusions or deform the composite with isotropic inclusion distribution to independently control the concentration ``in plane'' and perpendicular to it. 

(iii)~A layered structure based either on multiple semiconductor quantum wells~\cite{podolskiyGmachl} or on plasmonic (polar) materials~\cite{shvetsPRB}. 

We also anticipate the desired response from intrinsically anisotropic semi-metal crystals (Bi, and its alloys)~\cite{Bi}.

The work was partially supported by NSF grant DMR-0134736.


\begin{thebibliography}{}
\bibitem{losses} Here we restrict ourselves to the case of lossless LHM. Presence of finite losses, will inevitably limit the ``restoration ability'' of the LHM material, as it has been shown for $3D$ LHM lens in \cite{podolskiyResolut}.
\end{thebibliography}
\end{document}